# Identifying Clusters of Concepts in a Low Cohesive Class for Extract Class Refactoring Using Metrics Supplemented Agglomerative Clustering Technique

A. Ananda Rao[1] and K. Narendar Reddy[2]

[1] Professor of CSE and Principal, JNTUACE
Anantapur, Andhra Pradesh, India

[2] Associate Professor, Dept. of CSE, CVR College of Engineering
Hyderabad, India

**Abstract**
Object oriented software with low cohesive classes can increase maintenance cost. Low cohesive classes are likely to be introduced into the software during initial design due to deviation from design principles and during evolution due to software deterioration. Low cohesive class performs operations that should be done by two or more classes. The low cohesive classes need to be identified and refactored using extract class refactoring to improve the cohesion. In this regard, two aspects are involved; the first one is to identify the low cohesive classes and the second one is to identify the clusters of concepts in the low cohesive classes for extract class refactoring. In this paper, we propose metrics supplemented agglomerative clustering technique for covering the above two aspects. The proposed metrics are validated using Weyuker's properties. The approach is applied successfully on two examples and on a case study.

## 1. Introduction

The maintainability of object oriented software depends on the quality of software. Software quality is likely to get reduced due to deviation from design principles and due to software deterioration during evolution. The reduction in software quality can be attributed to the presence of bad smells. Low cohesive class is one of the bad smells. Object oriented software with low cohesive classes can increase maintenance cost. Defects (bad smells) in software cause the system to exhibit high complexity, improper behavior, and poor maintainability [1]. It is necessary to detect and correct the defects to make software maintainable. One of the ways to make object oriented software systems maintainable is refactoring. Techniques that reduce object oriented software complexity by incrementally improving the internal software quality without affecting the external behavior come under refactoring [2]. In the context of software under evolution, refactoring is used to improve the software quality. The improvement in the software quality is, in terms of, maintainability, complexity, reusability, efficiency, and extensibility [3].

In the literature active research is being carried out with respect to object oriented software refactoring. Considering the importance of low cohesive classes and refactoring, we have proposed an approach for identifying low cohesive classes and clusters of concepts in low cohesive classes for extract class refactoring. The low cohesive class indicates the presence of god class or divergent change bad smell. Divergent change bad smell is present in object oriented software whenever a class needs to be changed for different changes for different reasons in other classes [4]. The changes are propagated to the affected class due to rippling effects. The change propagations depend on strength of coupling between classes. Using strength of coupling between classes, our metric DOCMA(AR) (dependency oriented complexity metric for an artifact affected by ripples) can indicate the class affected by divergent change bad smell [5]. This bad smell indicates that the affected class has low cohesion and is a large class. According to Demeyer [6] the god class is low cohesive and memory consuming class. Object oriented software with low cohesive classes can increase maintenance cost. Low cohesive class performs operations that should be done by two or more classes. One of the design principles in object oriented approach is to design the class with "single minded" function [12]. The violation of this design principle results in large, complex, and low cohesive classes. The low cohesive class cohesion can be improved by splitting the class by extracting cohesive and independent groups of members addressing different functionalities using extract class refactoring [4]. A concept is a cluster (group) of class members addressing a single minded function. In this regard, two aspects are involved; the first one is to identify the low cohesive class and the second one is to identify clusters of concepts in low cohesive class which need to be refactored using extract class refactoring. In this paper, we propose an approach which is based on metrics supplemented hierarchical agglomerative clustering technique for covering the above two aspects. In this paper, "agglomerative clustering technique" means "agglomerative clustering algorithm".

Even though clustering techniques are mainly used in data mining, they are being applied successfully in software engineering. Clustering techniques can identify groups of similar entities [7],[8]. Clustering methods have very good potential to be used in software engineering [9] indicated by its use in software remodularisation[10]. Clustering techniques can identify groups of similar entities where in each group is conceptually different and these groups may address different functionalities.

The contributions of the proposed approach are the following:

- Metrics are proposed to supplement the agglomerative clustering technique (ACT) to handle the situations where some small clusters are formed.
- Identification of low cohesive classes.
- Identification of clusters of concepts in a low cohesive class using the approach which is based on metrics supplemented agglomerative clustering technique.

The organization of the paper is as follows. Section 2 presents the related work. Section 3 presents approach to clustering. The results of experimental cases are presented and discussed in section 4. The conclusions have been placed in section 5.

## 2. Related Work

The low cohesive classes can be caused by god class or divergent change bad smells. Lot of research has been done in identifying god classes. The main contributions are from [11], [1], [6]. According to Deymer [6] the god class is low cohesive and memory consuming class. Any change to the system may lead to this class. According to Trifu and Marinescu [11] god classes are "large, non-cohesive classes that have access to many foreign data" and they proposed a metrics-based method to identify. Tahvildari and Kontogiannis [1] proposed quality design heuristics and use a diagnosis algorithm based on coupling, complexity, and cohesion to identify design problems (flaws).

Lot of work has been done with respect to remodularising or partitioning or clustering large software modules. Some of them are [13], [14], [15]. In all of the above works remodularisation of software modules in a higher level (like package or file level) is proposed. We need clustering at class level.

Some of the works which focuses on software clustering at class level are: Simon et al. [16] suggested visualization techniques to identify extract class opportunities. Visualizing large classes can be difficult and make it difficult to identify clear clusters. De Lucia et al. [17] propose a methodology that uses structural and semantic metrics for identifying extract class refactoring opportunities. The semantic cohesion metric is based on the names of classes and entities which can be developer dependent hence, may change the results. In a recent work, Joshi and Joshi [18] uses concept lattice for identifying extract class refactoring opportunities. It is identified by the authors that for large systems the lattices can become very complex for the designer to identify problematic cases by inspecting the lattice visually. An algorithm [26] is proposed by the authors to find clusters based on similarity matrix. It is likely to consume more time to identify the clusters and it is threshold value dependant.

In most recent work, Marios Fokaefs et al [19] apply the agglomerative clustering algorithm for several threshold values (ranging from 0.1 to 0.9) and present all possible results to the user. Algorithm also identifies clusters of cohesive entities ranked according to their impact on the design of the whole system and presented to the designer. In contrary to [19], instead of presenting results at different thresholds, we compute clusters at particular threshold and supplement with metrics to merge small clusters with other clusters. In our approach, we supplement the agglomerative clustering technique (ACT) with metrics which can handle situations where ACT alone cannot give acceptable clusters in some situations. Our approach is not much dependent on the threshold value if it is not chosen either near 0 or near to 1. Our approach reduces much human intervention.

## 3. Proposed Approach

The aim of the proposed approach is to identify the low cohesive classes and clusters of concepts in low cohesive classes.

The proposed approach contains two steps:

**Step 1:**
Identify low cohesive classes using the metrics LCOM[22], TCC[23], and DOCMA(AR) [5]. LCOM and TCC metrics values can be used to identify the low cohesive class due to god class bad smell. Whereas, DOCMA(AR) metric value can be used to identify the low cohesive class due to divergent change bad smell.

**Step 2:**
The metrics supplemented agglomerative clustering technique is applied on low cohesive class which is found in step 1. During step 2, the clusters of concepts which need to be refactored are identified. During this step, firstly, similarities between the class members are calculated and then the agglomerative clustering algorithm (technique) is applied to find the clusters at a specified threshold. At the specified threshold there may be some small clusters which need to be merged, this is done with the help of proposed metrics. The agglomerative clustering technique is explained in the following section 3.1.

3.1 Agglomerative Clustering Technique (ACT)

The agglomerative clustering algorithm [8] (which is a hierarchical clustering algorithm) is used in this paper.
The Agglomerative Clustering Algorithm (Technique) is given below:

Step 1: Assign each entity (class member) to a single cluster.

Step 2: Repeat merging while the specified threshold value is not reached.
- Merge two closest clusters according to the considered merging criteria.

Step 3: Display the outcome of the algorithm as a hierarchy of clusters (Dendrogram).

The algorithm requires a threshold value for the similarity metric as a cut-off value. The clusters which are output from the algorithm are at the threshold (cut-off) value. The output hierarchy of the clusters is usually represented by a dendrogram. It has tree like structure. The leaves of the tree represent the individual (single) entities. During the merging process intermediate nodes are formed, they are actual clusters to be output based on cut-off value. The root is the final cluster which contains all the entities. The tree height can be represented using distance metric value or similarity metric value. We used similarity metric value in this paper.

In Hierarchical agglomerative clustering algorithm different linkage methods are available.
1. Single linkage   2. Complete linkage
3. Average linkage  4. Weighted linkage

According to Anquetil and Lethbridge [20] single linkage gives less coupled clusters, complete linkage favors more cohesive clusters, and average linkage gives clusters somewhere in-between the above two. In this paper we have used complete linkage method. The similarity metric we used is the Jaccard similarity metric. Anquetil and Lethbridge [20] indicate in their paper that jaccard distance metric produces good results in software remodularisation. To define the Jaccard similarity metric [24] between two class members we employ the notion of property set of class member. The property set for a method (PSet_$m_i$) is, the method itself and the methods and fields used (accessed/called) by that method. The property set for a field (PSet_$f_i$) is, the field itself and the methods using (accessing) it. These property sets are similar to dependency sets used by Simon et al[16].

**Similarity Matrix**: Similarity matrix (m **x** m) is constructed using the computed values of the above similarity based metrics. Where, m is number of members of the class. Class member means it can be a class method or field (variable).

The **Jaccard index**, also known as the **Jaccard similarity coefficient** (originally coined *coefficient de communauté* by Paul Jaccard [24]), is used for comparing the similarity and diversity of entities (sample sets).

Based on defined property sets we calculate the Jaccard similarity metric between two class members A and B as follows:

$$JSimM(A,B) = \frac{|A \cap B|}{|A \cup B|}$$

Similarity between two entities depends on the properties which are shared [25].

3.2 Proposed Metrics to Supplement Agglomerative Clustering Technique

The clusters identified at cut-off value may give clusters with very few members (may be single member clusters),

hence we need to merge them with other clusters. For this purpose we have proposed some metrics. The metrics are given below.

**CIM_V** - Cluster Identification Metric for Variable = ratio of number of methods within the cluster that reference the variable under consideration to the total number of methods in the cluster

**$CIM_{VR}\_M$** - Cluster Identification Metric for Method (Based on Variable References) = ratio of number of variables within the cluster referred by the method to the total number of variables in the cluster.

**$CIM_C\_M$** - Cluster Identification Metric for Method (Based on Method Calls) = ratio of number of methods within the cluster called by method under consideration to the total number of methods in the cluster

**$CIM_I\_M$** - Cluster Identification Metric for Method (Based on Method Invocations) = ratio of number of methods within the cluster that invoked method under consideration to the total number of methods in the cluster.

**Contexts for applying the proposed metrics:**

The contexts in which the proposed metrics can be applied are given in Figures 1, 2, 3, 4, and 5. The cluster to be merged with other cluster is indicated by C1.

The contexts are:
Context 1: Only one variable (field) in C1 (Figure 1).
Context 2: Only one method in C1 (Figure 2).
Context 3: Only variables (two or more variables).
Context 4: Only methods (two or more methods)
Context 5: Methods and variables. (At least one method and one variable)

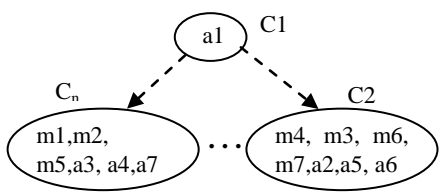

Fig. 1 Context1 for applying proposed metrics

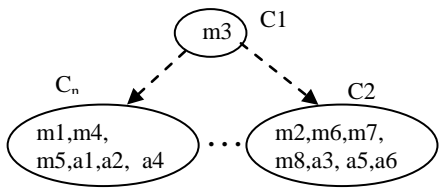

Fig. 2 Context2 for applying proposed metrics

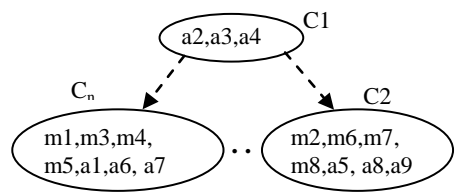

Fig. 3 Context3 for applying proposed metrics

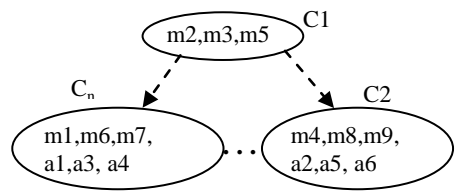

Fig. 4 Context4 for applying proposed metrics

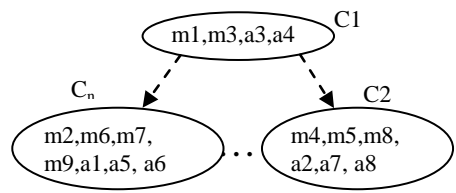

Fig. 5 Context5 for applying proposed metrics

The proposed metrics can be applied for all the five contexts. In all the contexts (Figures 1 to 5) proposed metrics can be used in merging the element(s) from cluster 1 (C1) with cluster 2 (C2) or cluster 3 (C3) or cluster n ($C_n$).

### 3.2.1 Validation using Weyuker's Properties

The proposed metrics are validated using the Weyuker's properties [21]. The validation results are given in Table 1.

Table 1: Validation results

|  | P1 | P2 | P3 | P4 | P5 | P6 | P7 | P8 | P9 |
|---|---|---|---|---|---|---|---|---|---|
| **CIM_V** | Y | Y | Y | Y | Y | Y | NA | Y | Y |
| **$CIM_{VR}\_M$** | Y | Y | Y | Y | Y | Y | NA | Y | Y |
| **$CIM_C\_M$** | Y | Y | Y | Y | Y | Y | NA | Y | Y |
| **$CIM_I\_M$** | Y | Y | Y | Y | Y | Y | NA | Y | Y |

Only eight properties are considered for validation. The seventh property is not considered for validation. Seventh property specifies that the permutation of elements within measured entity can change the metric value. This property is not suitable for OO systems, since the order of methods declaration inside a class does not change the order in which they are executed. Cherniavsky and Smith [27] suggest that this property is not appropriate for OOD metrics.

For contexts 1, 2, and 3 all the proposed metrics satisfied all the eight properties which are applicable to object oriented systems. In case of context 4 the metrics $CIM_C\_M$ and $CIM_I\_M$ did not satisfy the properties 5 and 9. In case of context 5, all the metrics did not satisfy properties 5 and 9. However, at least, 6 properties are satisfied out of 8 properties in all the contexts. The validation results indicate the suitability of metrics for the purpose for which they are proposed.

## 4. Experimental Results

**Example 1:**

The directed graph for the members of the class considered in this example is given in Figure 6. In the graph, the nodes represent methods and variables (fields) and the edges indicate that a dependency exists between two class members.

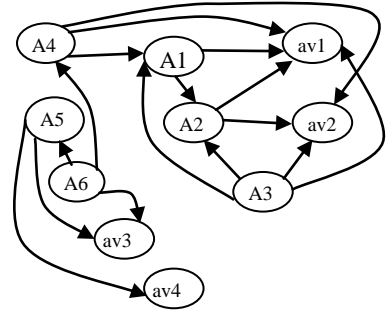

Fig. 6 Directed graph for the members of class in example 1

The class has six methods and four variables (fields).

The methods are: A1, A2, A3, A4, A5, A6
The variables are: av1, av2, av3, av4

The Jaccard similarity metric is used for finding the similarity between the class members. Similarity matrix for the members of the class given in example 1 is shown in Table 2. The hierarchical agglomerative clustering algorithm is applied on the similarity matrix (Table 2). Figure 7 shows the dendrogram produced by the algorithm.

Table 2: Similarity matrix for the class members shown in Figure 6

|     | A1   | A2   | A3   | A4   | A5   | A6   | av1 | av2 | av3  | av4 |
|-----|------|------|------|------|------|------|-----|-----|------|-----|
| A1  | 1    |      |      |      |      |      |     |     |      |     |
| A2  | 0.5  | 1    |      |      |      |      |     |     |      |     |
| A3  | 0.6  | 0.6  | 1    |      |      |      |     |     |      |     |
| A4  | 0.4  | 0.4  | 0.5  | 1    |      |      |     |     |      |     |
| A5  | 0    | 0    | 0    | 0    | 1    |      |     |     |      |     |
| A6  | 0    | 0    | 0    | 0.14 | 0.4  | 1    |     |     |      |     |
| av1 | 0.5  | 0.33 | 0.67 | 0.5  | 0    | 0.12 | 1   |     |      |     |
| av2 | 0.17 | 0.4  | 0.5  | 0.33 | 0    | 0.14 | 0.5 | 1   |      |     |
| av3 | 0    | 0    | 0    | 0    | 0.5  | 0.75 | 0   | 0   | 1    |     |
| av4 | 0    | 0    | 0    | 0    | 0.67 | 0.2  | 0   | 0   | 0.25 | 1   |

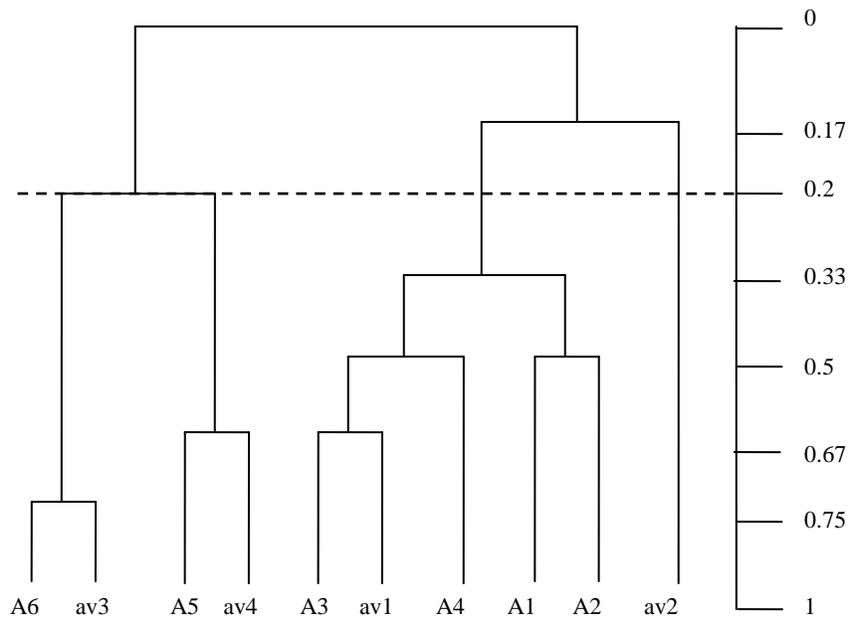

Fig. 7 Dendrogram showing the clusters

Clusters Identification at a given threshold value:

Consider the threshold value of 0.2. At this threshold, three clusters are produced, they are:
G1= {A5, A6, av3, av4}
G2= {av2}
G3= {A1, A2, A3, A4, av1}

Since G2 contain only one member (variable member) it needs to be merged with other cluster. This represents context 1 (Figure 1). For this purpose, **CIM_V** metric is computed.

**CIM_V** (av2, G1) = 0/2=0,
**CIM_V** (av2, G3) = ¾ = 0.75
Since **CIM_V** (av2) is more with respect to G3 when compared to G1, hence G2 (av2) is merged with G3.

After merging two clusters are formed. They are:
G1= { A5, A6, av3, av4},
G3= { A1, A2, A3, A4, av1, av2}

**Example 2:**

The directed graph for the members of the class considered in this example is given in Figure 8.

The class has seven methods and two variables (fields).
The methods are: A1, A2, A3, A4, A5, A6, A7
The variables are: av1, av2

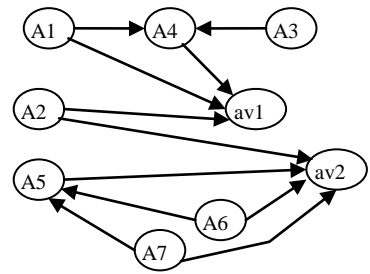

Fig. 8 Directed graph for the members of class in example 2

The similarity matrix is constructed using Jaccard similarity metric values for the class members and the hierarchical agglomerative clustering algorithm is applied on the similarity matrix. The clusters formed at 0.5 threshold value are:

G1= {A1, A4, A3}
G2= {av1}
G3= {A2}
G4= {A5, A6, A7}

G5= {av2}

Since single member clusters are there, they need to be merged. Merging G2 or G5 with other cluster represents context 1. Whereas, merging G3 with other cluster represents context 2 (Figure 2).

For context 1 the metric **CIM_V** is computed.

**CIM_V** (av1, G1) = 2/3=0.66
**CIM_V** (av1, G4) = 0/3=0

Since **CIM_V** metric value for av1 (G2) with respect to G1 is more when compared to G4, hence G2 is merged with G1. Similarly, av2 (G5) is merged with G4.

Since merging G3 (A2) with other cluster represents context 2, the metrics $CIM_{VR}\_M$, $CIM_C\_M$, and $CIM_I\_M$ are computed.

$CIM_{VR}\_M$ (A2, G1) = 1/1 = 1
$CIM_{VR}\_M$ (A2, G4) = 1/1 = 1
$CIM_C\_M$ (A2, G1) = 0/3 = 0
$CIM_C\_M$ (A2, G4) = 0/3 = 0
$CIM_I\_M$ (A2, G1) = 0/3 = 0
$CIM_I\_M$ (A2, G4) = 0/3 = 0

Since same metrics values indicate A2 (G3) can be placed either in G1 or G4.
The formed clusters after merging are:
G1= {A1, A4, A3, av1, A2}
G4= {A5, A6, A7, av2}

In situations where a member has equal similarity with two groups, (for example, in the above case) coupling with respect to other classes need to be computed before merging.

### Case Study : Bank Application

Bank application developed by students as part of their academic project is given as case study. The novice designer has a tendency to deviate from the design principles and may end up designing low cohesive classes. The bank application is developed in java. It contains the following classes:
CustomerAccount, Bank, ATMCard, Locker, DemandDraft, FixedDeposit, Employee, Manager, Report, Salary, DatabaseProxy. The proposed approach is applied on this case study to find low cohesive classes and clusters of concepts in low cohesive classes. The approach contains two steps.

Table 3: Similarity matrix for the members of class "CustomerAccount"

|     | M1 | M2 | M3 | M4 | M5 | M6 | M7 | M8 | M9 | M10 | M11 | V1 | V2 | V3 | V4 | V5 | V6 | V7 | V8 | V9 | V10 | V11 |
|-----|----|----|----|----|----|----|----|----|----|-----|-----|----|----|----|----|----|----|----|----|----|-----|-----|
| M1  | 1  | .5 | .5 | .43| .43| .1 | 0  | 0  | .1 | 0   | 0   | .17| .2 | .33| .25| .22| 0  | 0  | 0  | 0  | 0   | 0   |
| M2  |    | 1  | .7 | .67| .25| .1 | 0  | 0  | .1 | 0   | 0   | .27| .33| 0  | .25| .38| 0  | 0  | 0  | 0  | 0   | 0   |
| M3  |    |    | 1  | .67| .5 | .1 | 0  | 0  | .1 | 0   | 0   | .27| .33| 0  | .25| .38| 0  | 0  | 0  | 0  | 0   | 0   |
| M4  |    |    |    | 1  | .33| .13| 0  | 0  | .13| 0   | 0   | .2 | .25| 0  | 0  | .29| 0  | 0  | 0  | 0  | 0   | 0   |
| M5  |    |    |    |    | 1  | .13| 0  | 0  | .13| 0   | 0   | .33| .43| 0  | .14| .13| 0  | 0  | 0  | 0  | 0   | 0   |
| M6  |    |    |    |    |    | 1  | .5 | .5 | .11| 0   | 0   | .18| 0  | 0  | 0  | 0  | .29| .29| .29| 0  | 0   | 0   |
| M7  |    |    |    |    |    |    | 1  | .6 | 0  | 0   | 0   | 0  | 0  | 0  | 0  | 0  | 33 | .33| .33| 0  | 0   | 0   |
| M8  |    |    |    |    |    |    |    | 1  | 0  | 0   | 0   | 0  | 0  | 0  | 0  | 0  | .33| .4 | .33| 0  | 0   | 0   |
| M9  |    |    |    |    |    |    |    |    | 1  | .5  | .5  | .18| 0  | 0  | 0  | 0  | 0  | 0  | 0  | .29| .29 | .29 |
| M10 |    |    |    |    |    |    |    |    |    | 1   | .6  | 0  | 0  | 0  | 0  | 0  | 0  | 0  | 0  | .33| .33 | .33 |
| M11 |    |    |    |    |    |    |    |    |    |     | 1   | 0  | 0  | 0  | 0  | 0  | 0  | 0  | 0  | .33| .33 | .33 |
| V1  |    |    |    |    |    |    |    |    |    |     |     | 1  | .56| .11| .33| .44| .1 | 0  | 0  | .1 | 0   | 0   |
| V2  |    |    |    |    |    |    |    |    |    |     |     |    | 1  | .14| .43| .57| 0  | 0  | 0  | 0  | 0   | 0   |
| V3  |    |    |    |    |    |    |    |    |    |     |     |    |    | 1  | .2 | .17| 0  | 0  | 0  | 0  | 0   | 0   |
| V4  |    |    |    |    |    |    |    |    |    |     |     |    |    |    | 1  | .5 | 0  | 0  | 0  | 0  | 0   | 0   |
| V5  |    |    |    |    |    |    |    |    |    |     |     |    |    |    |    | 1  | 0  | 0  | 0  | 0  | 0   | 0   |
| V6  |    |    |    |    |    |    |    |    |    |     |     |    |    |    |    |    | 1  | .6 | .6 | 0  | 0   | 0   |
| V7  |    |    |    |    |    |    |    |    |    |     |     |    |    |    |    |    |    | 1  | .6 | 0  | 0   | 0   |
| V8  |    |    |    |    |    |    |    |    |    |     |     |    |    |    |    |    |    |    | 1  | 0  | 0   | 0   |
| V9  |    |    |    |    |    |    |    |    |    |     |     |    |    |    |    |    |    |    |    | 1  | .6  | .6  |
| V10 |    |    |    |    |    |    |    |    |    |     |     |    |    |    |    |    |    |    |    |    | 1   | .6  |
| V11 |    |    |    |    |    |    |    |    |    |     |     |    |    |    |    |    |    |    |    |    |     | 1   |

M1 – open(), M2 – deposit(), M3 – withdraw(), M4 – display(), M5 – close(), M6 – addCust(), M7 – updateAddr(), M8 – displayAddr(), M9 – apprLoan(), M10- repay(), M11-closeloan(), V1 – CustName, V2 – AcNo, V3 – AcType , V4 – Amount, V5 – Balance, V6 – CustId, V7 – PermAddr , V8 – CommnAddr, V9 – LoanNo, V10- LoanType, V11- LoanAmnt

Step 1:
During this step LCOM and TCC metrics are computed for the different classes. The CustomerAccount class has LCOM metric value of 5 and TCC metric value of 0.49. The high LCOM value and low TCC value indicate that the CustomerAccount is a low cohesive class.

Step 2:
The similarity matrix is constructed using Jaccard similarity metric values for the class members of the class "CustomerAccount" and it is shown in Table 3. The hierarchical agglomerative clustering algorithm is applied on the similarity matrix given in Table 3.

The clusters identified at the threshold value of 0.2 are:

G1 = M1, M2, M3, M4, M5
G2 = V1, V2, V4, V5
G3 = V3
G4 = M6, M7, M8, V6, V7, V8
G5 = M9, M10, M11, V9, V10, V11

Since G3 contains single variable and G2 contains only variables, they need to be merged with other groups. They represent contexts 1 and 3 respectively. Hence the metric **CIM_V** is computed to find the clusters with which G2, G3 can be merged.

Finding the cluster for G3(V3) merging:

**CIM_V** (V3) with respect to G1 = 1/5 = 0.2
**CIM_V** (V3) with respect to G4 = 0
**CIM_V** (V3) with respect to G5 = 0
Since **CIM_V** (V3) is high with respect to G1, hence V3 (G3) is merged with G1.

Finding the cluster for G2 merging:

**CIM_V** (V1) with respect to G1 = 5/5 = 1

**CIM_V** (V1) with respect to G4 = 1/3 =0.33
**CIM_V** (V1) with respect to G5 = 1/ 3 =0.33

**CIM_V** (V1) with respect to G1 is high when compared to G4 and G5 and **CIM_V** (V2, V4, V5) with respect to G1 is more when compared to G4 and G5, hence merge G2 with G1. After merging three clusters are formed, they are:

G1 = M1, M2, M3, M4, M5, V1, V2, V4, V5, V3
G4 = M6, M7, M8, V6, V7, V8
G5 = M9, M10, M11, V9, V10, V11

4.1 Observations

The LCOM [22] and TCC [23] are computed for the classes considered in the examples and case study before and after refactoring and presented in Table 4. The high LCOM and low TCC values indicate low cohesive classes, whereas low LCOM and high TCC values indicate high cohesive classes. The values in Table 4 indicate the improvement in cohesion due to refactoring in two experimental examples and in one case study.

The high value of LCOM and low value of TCC for the CustomerAccount class of bank application indicate the low cohesive class. The Agglomerative Clustering Technique identified three clusters of concepts for CustomerAccount class. These three clusters are refactored into three classes by using extract class refactoring. The extracted classes are: Account, Customer, and Loan. The 0 (zero) LCOM value and TCC value of 1 for Account (cluster1), Customer (cluster2), and Loan (cluster3) indicate high cohesive classes. Hence our approach could identify the low cohesive class, and the clusters to be refactored. The increase in cohesion after refactoring, indicate the effectiveness of the approach in identifying proper clusters for refactoring.

Table 4.  LCOM and TCC values before and after refactoring

|  | Before Refactoring | | After Refactoring | | | | | |
|---|---|---|---|---|---|---|---|---|
|  |  |  | Cluster1 | | Cluster2 | | Cluster3 | |
|  | LCOM | TCC | LCOM | TCC | LCOM | TCC | LCOM | TCC |
| Example 1 | 7 | .47 | 0 | 1 | 0 | 1 | - | - |
| Example 2 | 9 | .43 | 0 | .5 | 0 | 1 | - | - |
| Case Study | 5 | .49 | 0 | 1 | 0 | 1 | 0 | 1 |

The proposed approach is useful because of the following reasons:

- It handles the situation where Agglomerative clustering technique (ACT) alone cannot identify proper clusters. Presenting to the user clusters at different thresholds may not solve the problem. In some situations, clusters at any threshold other than maximum (which leads to a single cluster) may contain clusters with single members or small clusters with one member or clusters with only variables. In that situation ACT does not guide us how to merge those small clusters. Hence, proposed metrics will help in those situations.
- In our approach, we need to specify one threshold value somewhere in between 0 and 1, which is neither near to 0 nor near to 1. Merge small clusters with the help of proposed metrics. The rules (in what situations the groups should be merged) can be specified to the tool, which guide the merging process using the proposed metrics. Hence, they reduce the human intervention to study the clusters at different thresholds and decide.
- When a single member is tried to be merged, if metrics values are equal with respect to two or more clusters, it can be merged by considering coupling with respect to other classes.

## 5. Conclusions

In this paper, we proposed an approach for identifying low cohesive classes and clusters of concepts in low cohesive classes potential for extract class refactoring. Proposed approach consists of two steps. In step 1, low cohesive classes are identified. In step 2, the clusters of concepts in low cohesive classes are identified for extract class refactoring. The proposed approach is based on metrics supplemented agglomerative clustering technique. Agglomerative Clustering Technique is based on the Jaccard similarity metric values between class members. Metrics which are used to supplement agglomerative clustering technique are newly proposed. The metrics are validated using Weyuker's properties. The approach is applied on two examples and on academic software developed by students. In the two examples and case study our approach could find low cohesive classes and clusters of concepts to be refactored. The low cohesive class identified by our approach in the bank application has high LCOM and low TCC value. High LCOM and low TCC values indicate low cohesive class. The clusters identified by our approach are refactored using extract class refactoring. After refactoring the LCOM metric value is decreased whereas TCC metric value is increased. These values indicate increase in cohesion due to refactoring the clusters of concepts into new classes. The increase in cohesion after refactoring, indicate the effectiveness of the approach in identifying proper clusters for refactoring. Hence, our approach could effectively identify low cohesive classes and clusters of concepts to be refactored.

## References

[1] L. Tahvildari and K. Kontogiannis. "A Metric-Based Approach to Enhance Design Quality Through Meta-Pattern Transformations" In Proceedings of the 7th European Conference on Software Maintenance and Reengineering, March 26-28, 2003, pp. 183–192.

[2] W .F. Opdyke, "Refactoring : A Program Restructuring Aid in Designing Object-Oriented Application Frameworks", PhD thesis, Univ. of Illinois at Urbana Champaign, 1992.

[3] Tom Mens and Tom Tourwe. "A Survey of Software Refactoring", IEEE Transactions on Software Engineering. Volume 30, Number 2, 2004, pp 126-139.

[4] M. Fowler, K. Beck, J. Brant, W. Opdyke, and D. Roberts. Refactoring Improving the Design of Existing Code, Boston, MA, AddisonWesley, 1999.

[5] K.Narendar Reddy and A.Ananda Rao, "Dependency Oriented Complexity Metrics to Detect Rippling Related Design Defects", ACM SIGSOFT Software Engineering Notes, Volume 34, Number 4, July 2009.

[6] S. Demeyer, S. Ducasse, and O. M. Nierstrasz, Object-Oriented Reengineering Patterns, Morgan Kaufman Publishers, 2002.

[7] P. N. Tan, M. Steinbach, and V. Kumar, Introduction to Data Mining, Addison-Wesley, 2005.

[8] Jiawei Han and M. Kamber, Data Mining Concepts and Techniques, Morgan Kaufmann Publishers, 2005.

[9] V. Tzerpos and R. C. Holt, "Software Botryology: Automatic Clustering of Software Systems", In Proceedings of the International Workshop on Large-Scale Software Composition, 1998.

[10] T. A.Wiggerts, "Using Clustering Algorithms in Legacy Systems Remodularization", In WCRE '97: Proceedings of the 4th Working Conference on Reverse Engineering, 1997.

[11] A. Trifu and R. Marinescu, " Diagnosing Design Problems in Object Oriented Systems", In Proceedings of the 12th Working Conference on Reverse Engineering, 2005.

[12] Roger S. Pressman, Software Engineering A Practitioner's Approach, Sixth Edition, McGraw-Hill Int'l Edition, 2005.

[13] S. Mancoridis, B. S. Mitchell, C. Rorres, Y. Chen, and E. R.Gansner, "Using Automatic Clustering to Produce High-Level System Organizations of Source Code", In Proceedings of the 6th International Workshop on Program Comprehension, 1998 ,pp. 45–52.

[14] D. Doval, S. Mancoridis, and B. S. Mitchell "Automatic Clustering of Software Systems Using a Genetic Algorithm",. In Proceedings of the 5th International Conference on Software Tools and Engineering Practice, 30 August - 2 September 1999.

[15] K. Sartipi and K. Kontogiannis, "Component Clustering Based on Maximal Association", In Proceedings of the IEEE Working Conference on Reverse Engineering, October 2001.

[16] F. Simon, F. Steinbruckner, and C. Lewrentz, "Metrics Based Refactoring", In Proceedings of the 5th European Conference on Software Maintenance and Reengineering, 2011, pp.30–38.

[17] A. D. Lucia, R. Oliveto, and L. Vorraro, "Using Structural and Semantic Metrics to Improve Class Cohesion", In 24th IEEE International Conference on Software Maintenance, 2008.

[18] P. Joshi and R. K. Joshi, "Concept Analysis for Class Cohesion", In European Conference on Software Maintenance and Reengineering, March 24-27 2009, pp. 237–240.

[19] Marios Fokaefs, Nikolaos Tsantalis, and Alexander Chatzigeorgiou,"Decomposing Object-Oriented Class Modules Using an Agglomerative Clustering Technique", In Proc. ICSM, 2009, pp. 93-101.

[20] N. Anquetil and T. Lethbridge, "Experiments with Clustering as a Software Remodularization Method", In WCRE '99: Proceedings of the 6th Working Conference on Reverse Engineering, 1999.

[21] E.J.Weyuker, "Evaluating Software Complexity Measures", IEEE Transactions on Software Engineering, Vol. 14, No. 9, 1988, pp. 1357-1365.

[22] S.R.Chidamber and C.F.Kemerer, "A Metrics Suite for Object Oriented Design", IEEE Transactions on Software Engineering, Vol. 20, No. 6, 1994, pp. 476-493.

[23] J.M.Bieman and B.K.Kang, "Cohesion and Reuse in an Object-Oriented System", In Proc. ACM Symp. Software Reusability (SSR'94), 1995, pp. 259-262.

[24] Paul Jaccard . Étude comparative de la distribution florale dans une portion des Alpes et des Jura. Bulletin de la Société Vaudoise des Sciences Naturelles 37, 1901, pp. 547–579.

[25] Mario Bunge. Treatise on Basic Philosophy, Vol 3: Ontology I, The Furniture of the World, ordrecht_Holland, D.Reidel Publishing Company, 1977.

[26] K.Narendar Reddy and A.Ananda Rao, "Poster paper: Similarity Based Metrics for Performing Extract Class Refactoring for the Class Affected by Ripples", In ISEC(3[rd] India Software Engg. Conference), Feb 25-27, 2010, Mysore, India.

[27] J.C.Cherniavsky and C.H.Smith, " On Weyuker's Axioms for Software Complexity Measures", IEEE Transactions on Software Engineering, Volume 17, 1991, pp 636-638.

**Prof. Ananda Rao Akepogu** received B.Sc.(M.P.C) degree from Silver Jubilee Govt. College, SV University, Andhra Pradesh, India. He received B.Tech. degree in Computer Science & Engineering and M.Tech. degree in A.I & Robotics from University of Hyderabad, India. He received Ph.D. from Indian Institute of Technology, Madras, India. He is Professor of Computer Science & Engineering and Principal of JNTUA College of Engineering, Anantapur, India. Prof. Ananda Rao published more than fifty research papers in international journals, conferences and authored three books. His main research interests include software engineering and data mining.

**Narendar Reddy K** is pursuing Ph.D. in Computer Science & Engineering from JNTUA, Anantapur, India and he received his M.Tech. in Computer Science & Engineering from the same University. He received Bachelor's degree in Computer Science & Engineering (AMIE(CSE)) from Institution of Engineers, Calcutta, India. Currently he is working as Associate Professor of Computer Science & Engineering at CVR College of Engineering, Hyderabad, India. His main research interests include Object oriented software design, software metrics, refactoring, and software testing. He is a member of IEEE, ACM, AMIE(I), and IAENG.